\documentclass[journal=jacsat,manuscript=article,layout=twocolumn]{achemso}
\setkeys{acs}{articletitle=true}

\usepackage[version=3]{mhchem} 
\usepackage[T1]{fontenc}       
\usepackage{amsmath}
\usepackage{amssymb}
\usepackage{graphicx}
\usepackage{afterpage}
\usepackage{amsfonts}
\usepackage{textcomp}

\title{Reading the orbital angular momentum of light using plasmonic nanoantennas}

\author{R.~M.~Kerber}
\email{r.kerber@wwu.de} \affiliation{Institut f\"{u}r
Festk\"{o}rpertheorie, Universit\"{a}t M\"{u}nster, 48149
M\"{u}nster, Germany}

\author{J.~M.~Fitzgerald}
\affiliation{Department of Physics, The Blackett Laboratory, Imperial College London, South Kensington Campus, London SW7 2AZ, United Kingdom}

\author{D.~E.~Reiter}
\affiliation{Institut f\"{u}r Festk\"{o}rpertheorie, Universit\"{a}t
M\"{u}nster, 48149 M\"{u}nster, Germany}
\alsoaffiliation{Department of Physics, The Blackett Laboratory, Imperial College London, South Kensington Campus, London SW7 2AZ, United Kingdom}

\author{S.~S.~Oh}
\affiliation{Department of Physics, The Blackett Laboratory, Imperial College London, South Kensington Campus, London SW7 2AZ, United Kingdom}

\author{O.~Hess}
\email{o.hess@imperial.ac.uk} \affiliation{Department of Physics, The Blackett Laboratory, Imperial College London, South Kensington Campus, London SW7 2AZ, United Kingdom}

\begin{document}

\twocolumn[
\begin{@twocolumnfalse}
\begin{abstract}
Orbital angular momentum of light has recently been recognized as a new degree of freedom to encode information in quantum communication using light pulses. Methods to extract this information include reversing the process by which such twisted light was created in the first place or interference with other beams. Here, we propose an alternative new way to directly read out the extra information encoded in twisted light using plasmonic nanoantennas by converting the information about the orbital angular momentum of light into spectral information using bright and dark modes. Exemplarily considering rotation-symmetrical nanorod nanoantennas we show that their scattering cross-section is sensitive to the value of the orbital angular momentum combined with the polarisation of an incident twisted light beam. Explaining the twist-dependence of the excited modes with a new analytical model our results pave the way to twisted light nanoplasmonics, which is of central importance for future on-chip communication using orbital angular momentum of light.
\end{abstract}

\end{@twocolumnfalse}
]

\begin{figure}
\includegraphics[width=1\columnwidth]{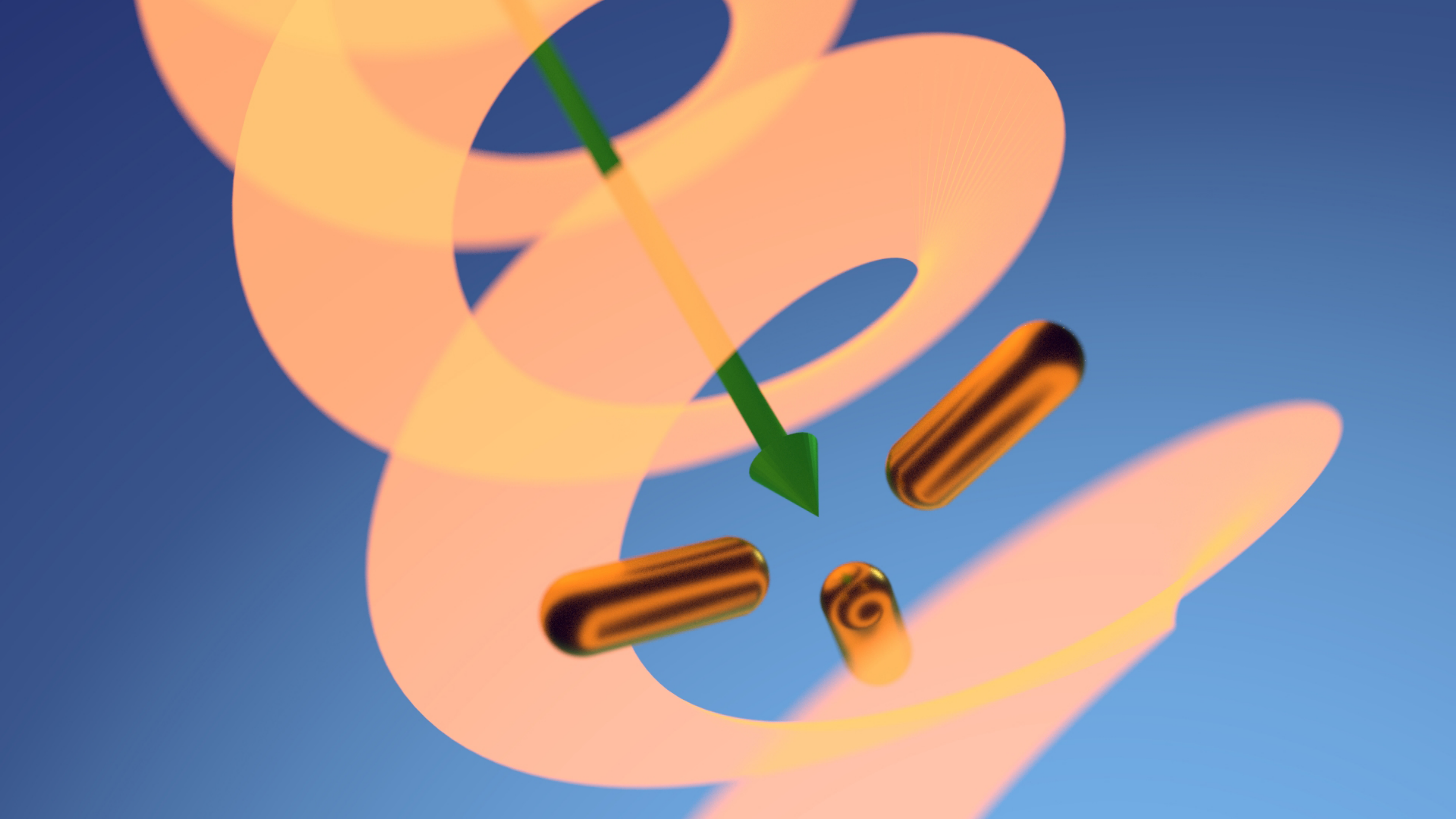}
\end{figure}
\section{Introduction}
Twisted light , i.e. light carrying orbital angular momentum (OAM) \cite{andrews2008structured}, has recently emerged as a new way to encode information into the phase of a light beam \cite{bozinovic2013terabit,ren2016chip,mirhosseini2015high,wang2012terabit}. Moreover, the OAM of twisted light can also be exploited in areas ranging from unusual transitions in semiconductors\cite{quinteiro2009theory,shigematsu2016coherent} to applications as tweezers in biology\cite{padgett2011tweezers,woerdemann2013advanced}. Having generated a twisted light beam \cite{karimi2014generating,yu2011light,pu2015catenary,genevet2012holographic,coles2013chiral,williams2013optical}, the information about the OAM can be retrieved by converting the twisted light back to plane waves \cite{mirhosseini2013efficient,genevet2012holographic}, wiping out the OAM of the twisted light beam. Another method is to use interference with other beams and detecting the resulting interference pattern \cite{karimi2014generating,wang2012terabit}. Here, we propose an alternative approach to read out the OAM of light using plasmonic nanoantennas and show that the OAM can be converted into spectrally sensitive information. An advantage of our approach is that well established spectral measurement technique can be used for detection rather than spatial measurements.
For twisted light, one has to keep in mind that in addition to the OAM such a beam can also have a certain circular polarisation, which in quantum optics is denoted in terms of spin angular momentum (SAM) in which information can also be encoded. Depending on the combination of OAM $\ell$ and handedness of polarisation $\sigma$, the twisted light beam can be categorised into two distinct classes, which we shall call the parallel class for $\ell$ and $\sigma$ having the same sign and the antiparallel class with $\ell$ and $\sigma$ having opposite signs \cite{quinteiro2015formulation}. In this paper, we will show that the resonance frequency of a rotationally arranged nanoantenna is sensitive to the class of twisted light and to the absolute value of the OAM $|\ell|$. The resonances which are excited by the twisted light can be identified with bright and dark modes of plane waves \cite{nordlander2004plasmon}. In other words, twisted light can also be used to excite a dark mode, in addition to already existing possibilities like focused electron beam \cite{chu2008probing,schmidt2012dark}, far-field illumination techniques with spatially inhomogeneous fields \cite{huang2010mode}, evanescent excitation \cite{yang2010plasmon}, non-normal excitation \cite{hao2008shedding,zhou2011tunable} or radially or azimuthally polarized light \cite{volpe2009controlling,mojarad2009tailoring,sancho2012dark,gomez2013dark}. We further establish an analytical model to explain the number of modes that are excited by twisted light and discuss the degeneracy of different modes. This provides a direct way to explain how a nanoantenna reacts to twisted light.

\section{Theoretical background}

\begin{figure}
\includegraphics[width=1\columnwidth]{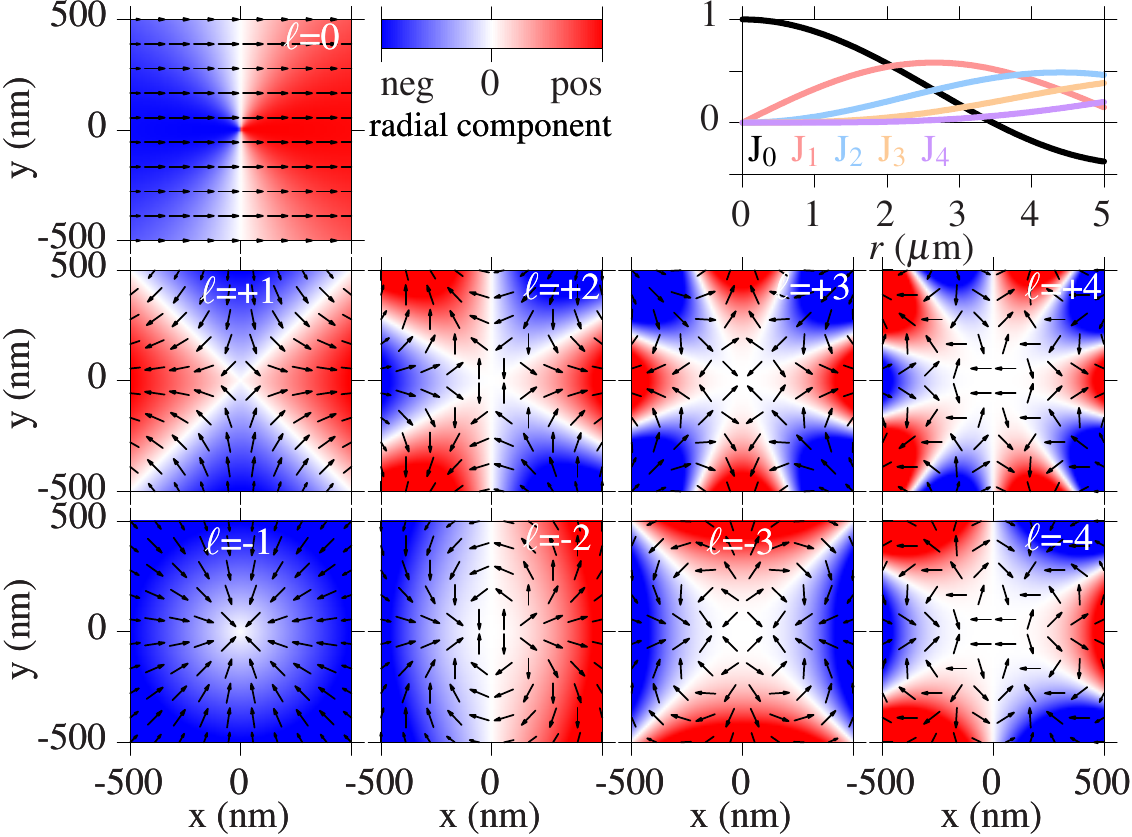}
\caption{Electric field patterns of twisted light with handedness of polarisation $\sigma=+1$. The OAM in the top left panel is $\ell=0$, corresponding to circularly polarized light The middle row shows twisted light in the parallel class with $\ell=1 \dots,4$ and the bottom row represents the anti-parallel class with $\ell=-1\ldots-4$. All fields are in the $z=0$ plane at $t=0$. The arrows show the direction of the electric field vectors and the background color indicates the strength of the radial component. Top right: Bessel functions $J_0 \ldots J_4$ as function of the distance $r$ from the beam axis. 
\label{fig:fieldpattern}}
\end{figure}

Let us first revisit the properties of a twisted light beam. The most important property is the additional phase describing the OAM, which also leads to the formation of a vortex or phase singularity at the beam axis. The radial profile of a twisted light beam can be described mathematically either in terms of a Laguerre-Gaussian \cite{loudon2003theory} or Bessel function \cite{jauregui2004rotational}.  Here we shall consider Bessel beams, because they are exact solutions of Maxwell's equations \cite{saleh2007fundamentals}. For Bessel beams, the electric field in cylindrical coordinates $\{r,\varphi,z\}$ with $\mathbf{E}=E_r \vec{e}_{r} + E_{\varphi} \vec{e}_{\varphi}+ E_z \vec{e}_{z}$ of a monochromatic Bessel beam with wavelength $\lambda$ and a propagation direction along the $z$-axis can be expressed as follows \cite{jauregui2004rotational,quinteiro2014light,quinteiro2015formulation}:
\begin{eqnarray}
\notag
E_r(\textbf{r},t) &=& E_0 J_{\ell}(q_r r) \times\\
\notag
&&\sin[(\omega t-q_z z)-(\ell+\sigma)\varphi], \\
\notag
E_{\varphi}(\textbf{r},t) &=& -\sigma  E_0 J_{\ell}(q_r r) \times\\
\notag
&&\cos[(\omega t-q_z z)-(\ell+\sigma)\varphi],\\
\notag
E_{z}(\textbf{r},t) &=& \sigma \frac{q_r}{q_z} E_0 J_{\ell+\sigma}(q_r r) \times\\
&&\cos[(\omega t-q_z z)-(\ell+\sigma)\varphi],
\end{eqnarray}
with the frequency $\omega=2\pi c/\lambda$, the velocity of light $c$, the wave vectors $q_z$ along the propagation direction and $q_r$ in the transversal plane with a fixed ratio of $q_r/q_z=0.1$. $J_{\ell}(q_r r)$ is the Bessel function as shown in the top right of Fig.~\ref{fig:fieldpattern}, which except for $\ell=0$ is zero at the origin reflecting the phase singularity. The OAM is encoded in $\ell$, which is also called topological charge, while $\sigma=\pm 1$ denotes the handedness of circular polarisation. Due to the symmetry, there is a degeneracy when the sign of both $\ell$ and $\sigma$ change simultaneously.  Therefore, it is sufficient to restrict ourselves in the following to $\sigma=+1$.

It is instructive to look at a plot of the transverse electric field of twisted light for different values of $\ell$ at $z=0$ presented in Fig.~\ref{fig:fieldpattern}. For the parameters we set $\lambda=900\,\text{nm}$ such that $q_r\approx 7 \times 10^{-4} \,\text{nm}^{-1}$. The field patterns for $\ell=0$ in the top left of Fig.~\ref{fig:fieldpattern} is analogous to the usual circularly (or even linearly) polarized light with the vectors of the electric field all pointing in the same direction. The middle row shows the electric field for $\ell=+1\ldots+4$, while the bottom row displays the field for the values $\ell=-1\ldots -4$. The complex field patterns show that the twisted light falls into two classes, the parallel class with $\text{sgn}(\ell)=\text{sgn}(\sigma)$ (middle row) and the antiparallel class with $\text{sgn}(\ell)\neq\text{sgn}(\sigma)$ (bottom row). Strikingly, even when evolving in time the field patterns do not evolve into each other \cite{quinteiro2015formulation}. 

We further specify the radial component of the electric field by the different background colours, where red (positive) refers to an outpointing electric field, while blue (negative) indicates an in-pointing electric field. This will become important for the interaction with the nanoantennas as discussed below because the radial component of the electric field points in the same direction of the antenna arms and determines mainly the resonance behaviour. We already note here that the symmetry of the radial component is given by $\ell+\sigma$, e.g. for $\ell=0$ (and $\sigma=+1$) there is one change between positive and negative values, while for $\ell=-1$ (and $\sigma=+1$) the field is rotationally symmetrical, e.g. purely radial as shown in Fig.~\ref{fig:fieldpattern}.

We chose a design consisting of a rotationally arranged nanoantenna comprising $N$ identical nanorods and in the folllowing show results for $N=2,3,6$ symmetrically arranged nanorods. Without loss of generality we assume twisted light to be incident on the nanoantenna from the top, aligning the symmetry axis with the propagation direction. Each rod has a length of $L=150$\,nm and a circular cross-section with a diameter of $D=40$\,nm, while the ends of the rods are rounded by hemispheres. The rods are positioned symmetrically about the $z=0$ plane. The gap region in the center has a diameter of $G=50$\,nm. Moreover, we assume the antennas to be made of gold and surrounded by air. To calculate the scattering cross-section of the nanoantennas, we use a boundary element method (BEM) \cite{hohenester2012mnpbem} with experimental data for the dielectric function of gold \cite{johnson1972optical}.

To gain a deeper understanding of the selectivity of the nanoantennas on the OAM of light, we here develop an analytical model using the `thin-wire' approximation of classical antenna theory \cite{orfanidis2002electromagnetic}.

For the model, we assume that each arm is a perfectly conducting wire of the length $L$, which also saves us from having to make the distinction between volume and surface currents which is necessary for antennas in the visible regime \cite{dorfmuller2010plasmonic}. The electric field radiating from a single wire can be calculated by its vector potential $\vec{A}$, where we assumed a harmonic time dependence \cite{orfanidis2002electromagnetic}. The retarded vector potential then reads 
\begin{eqnarray}
	\label{eq:vectorpotential}
	\vec{A}(\vec{r}) = \frac{\mu_0}{4\pi}\int d^{3}r' \frac{\vec{J}(\vec{r'})}{|\vec{r}-\vec{r'}|} e^{ik|\vec{r}-\vec{r'}|},
\end{eqnarray}
where $\vec{J}$ is the current density in the wire and $k$ is the wave vector of the radiation field. Fixing the antenna along the $x$-direction and assuming that it has no width in the $y$- and $z$-directions yields a current density $\vec{J}$ of the form
\begin{eqnarray}
	\label{eq:currentdensity}
	\vec{J}(\vec{r}) = I(x)\delta(y)\delta(z) \vec{e}_{x},
\end{eqnarray}
where $I(x)$ is the current in the wire. We assume that at the resonance wavelength $\lambda_r$ the current is of the form
\begin{eqnarray}
	\label{eq:current}
	I(x)=I_0 \cos(kx),
\end{eqnarray}
with $k=2\pi/L$ and the length of the antenna is $L=\lambda_r/2$. We note that this is an approximation, seeing that in the BEM the resonance frequency is not twice the antenna length, however, the model is well suited to explain qualitative features found in the numerical calculation. With the current the retarded vector potential $\vec{A}$ (Eq.~\ref{eq:vectorpotential}) can be evaluated \cite{orfanidis2002electromagnetic}. Using Lorenz gauge the scalar potential is given by $\Phi={\vec{\nabla} \cdot \vec{A}}/{(i\omega\mu\varepsilon)}$, where $\varepsilon=\varepsilon_0\varepsilon_r$ is the permittivity and $\mu=\mu_0\mu_r$ is the permeability. 
For air we take $\varepsilon_r=1$ and $\mu_r=1$. With this the electric field is determined via 
\begin{eqnarray}
\vec{E}(\vec{r},\omega)=\frac{-1}{i\omega\mu\varepsilon}[\vec{\nabla}(\vec{\nabla} \cdot \vec{A}(\vec{r},\omega)+k^2\vec{A}(\vec{r},\omega)].
\end{eqnarray}
The results are analytical formulas for the electric field of one arm of the antenna centered around $\vec{r}=0$ and pointing along the $x$-direction:
\begin{eqnarray}
\notag
E_x &=& \frac{i I_0}{4\pi c \varepsilon}[G_{-}+G_{+}]\\
\notag
E_y &=& \frac{-i I_0} {4\pi c \varepsilon} \frac{y}{y^2+z^2} \times\\
\notag
&&[(x-L/2)G_{-}+ (x+L/2) G_{+}]\\
\notag
E_z &=& \frac{-i I_0} {4\pi c \varepsilon} \frac{z}{y^2+z^2} \times\\
&&[(x-L/2)G_{-}+ (x+L/2) G_{+}].
\end{eqnarray}
$G_{\pm}$ is given by
\begin{eqnarray}
G_{\pm}=\frac{e^{ik\sqrt{(x\pm L/2)^2+y^2+z^2}}}{\sqrt{(x\pm L/2)^2+y^2+z^2}}.
\end{eqnarray}
The field $\vec{E}^{(n)}$ of the $n$-th arm of the antenna is obtained by rotating the wire by $\varphi_n={2\pi n}/{N}$ and then displacing it by the half of length and half of gap $(L+G)/2$. 

To calculate the total field $\vec{E}^{tot}$, which is generated by the nanoantenna excited by a twisted light beam with the OAM $\ell$ and the polarisation $\sigma$, we superimpose the resulting field of the $N$ arms and additionally include a phase factor $e^{i(\ell+\sigma)\varphi_n}$ to account for the phase of the twisted light beams. Then,
\begin{eqnarray}
	\vec{E}^{tot} = \sum_{n=0}^{N-1} \vec{E}^{(n)}e^{i(\ell+\sigma)\varphi_n},
\end{eqnarray}
which yields an analytical equation for the induced fields of the nanoantenna.

\section{Results and Discussion}

\begin{figure}
\centering
\includegraphics[width=1\columnwidth]{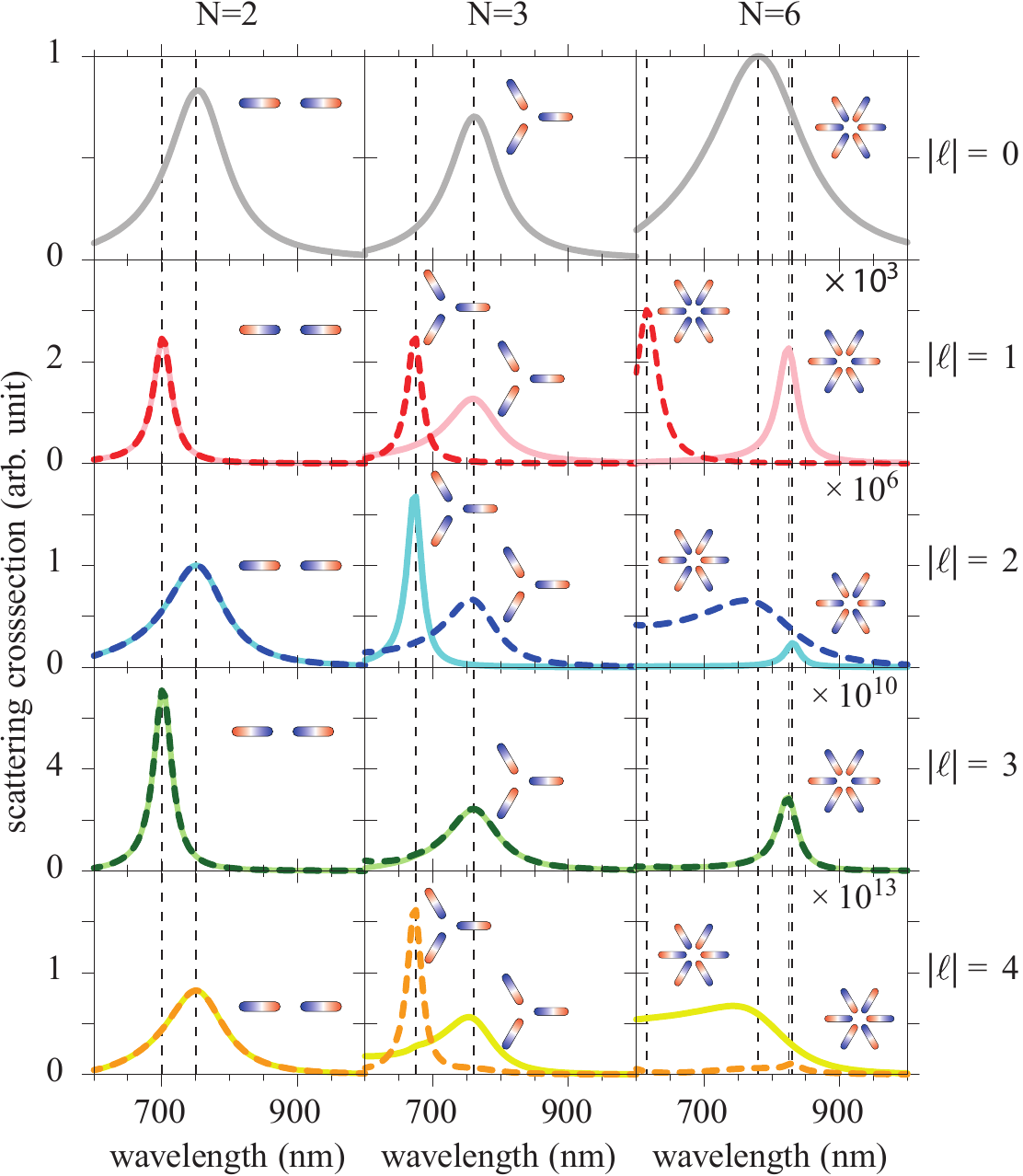}
\caption{Scattering cross-sections of a nanoantenna with $N=2$ (left), $N=3$ (middle) and $N=6$ (right) arms excited by twisted light calculated with the BEM. The rows corresponds to an OAM of the twisted light of $\ell=0$ to $|\ell|=4$ from top to bottom, respectively. The solid lines belong to the parallel class ($\text{sgn}(\ell)>0$) and the dashed lines to the antiparallel class ($\text{sgn}(\ell)<0$). The scaling of the cross section is denoted in the top left. The vertical dashed lines mark the resonance frequencies $\lambda_{d/b}$ excited by the twisted light. The insets show the surface charge distribution in the nanorods of the resonances with blue indicating a negative and red a positive surface charge.\label{fig:BEM}}
\end{figure}

First we discuss the results calculated with the BEM. The results are shown in Fig. \ref{fig:BEM} for different values of $\ell$ ranging from $|\ell|=0$ to $|\ell|=4$ from top to bottom, respectively, and for $N=2,3$ and $6$ from left to right. The scattering cross-section for the parallel class is marked by solid lines, while the antiparallel class is denoted by dashed lines. The resonances are marked by vertical black dashed lines. The insets show the corresponding surface charges of the resonances as calculated in the BEM with red indicating a positive and blue a negative charge. Due to the increasing beam waist the intensity drops for increasing OAM $|\ell|$, which is compensated by the factor noted in the top right of each row.

Let us first study the resonances in the scattering cross-section of the dimer antenna  with $N=2$ shown in the left column of Fig.~\ref{fig:BEM}. For $\ell=0$, the bright or bonding mode of the nanoantenna at $\lambda_b=750$~nm is excited. When we now increase $|\ell|$ we see that for even $|\ell|=2,4$ the bright resonance is excited, while for odd $|\ell|=1,3$ a different resonance at $\lambda_d=700$~nm comes up which corresponds to the dark or antibonding mode. The difference in the two modes can be explained using a hybridization model similar to two or more coupled harmonic oscillator resembling the surface charges \cite{nordlander2004plasmon,biagioni2012nan}. The strength of the splitting is very sensitive to the coupling between the two arms, which is mostly affected by the size of the gap $G$. The BEM allows us to calculate the surface charges as indicated in Fig.~\ref{fig:BEM}, showing that, indeed, in the bright mode the surface charges oscillate in-phase, while for the dark mode the surface charges oscillate out-of-phase. As for coupled harmonic oscillators we see that energetic position of the anti-bonding mode is above the one of the bonding mode \cite{nordlander2004plasmon,biagioni2012nan}. The relation between the phases of the surface charges and the twisted light can be determined by looking at the corresponding radial component of the incident beam. As shown in Fig.~\ref{fig:fieldpattern}, for even $|\ell|=0,2,4$, we see that there is a change between positive and negative radial components in the $x$-direction, while for odd $|\ell|=1,3$, the opposite lying radial components have the same direction. We further note that the dark resonance is narrower than the bright one. This is related to the fact that the bright mode corresponds to dipole radiation, while the dark mode has a quadrupolar radiation pattern. For $N=2$ we thus conclude that the dimer antenna is sensitive to the absolute value of OAM and that it is possible to distinguish between even and odd values of OAM. However, the dimer antenna does not distinguish between the two classes of twisted light.

If we now consider a trimer nanoantenna with $N=3$ as displayed in the middle column of Fig.~\ref{fig:BEM} we see that different resonances for the two distinct classes of twisted light are excited. For $\ell=0$ there is a resonance at $\lambda_b=760$~nm accompanied by a surface charge distribution where one rod has a positive charge pointing towards the middle while the other two have the opposite charge in the center. We call this mode again `bright'. For $|\ell|=1$ we see that in the parallel class with $\ell=+1$ (solid line) also the bright mode is excited, while in the antiparallel class with $\ell=-1$ (dashed line) a resonance at $\lambda_d=675$~nm is excited. For this dark resonance all surface charge distributions point synchronously either inwards or outwards. Indeed, for $N=3$ only the modes with these two frequencies can be excited when considering the phase combinations of surface charges in the three arms. For $|\ell|=2$ we see again a splitting between the classes, but the other way round to the case with $|\ell|=1$, i.e., the dark mode is excited by the parallel class, while the antiparallel class has a resonance at $\lambda_b$. If we now look at $|\ell|=3$, which corresponds to the number of arms of the nanoantenna, we find that both parallel and antiparallel classes are again degenerate and both excite the bright mode. 

The link between the resonant modes and the OAM of the excited light can be explained by considering the symmetry of the radial components of the twisted light (cf. Fig.~\ref{fig:fieldpattern}). As an example, we discuss the case of $|\ell|=2$. For parallel light, we find three axes where the radial component changes from inwards to outwards. When we consider the symmetry of the nanoantenna with $N=3$ it is clear that the arms always find themselves in a region with the same direction. On the other hand, in the antiparallel class with $\ell=-2$ there is only one axis. Hence the surface charge distribution in one arm always points in the opposite direction to the one in the other arms, corresponding to the bright mode. 

The higher the number of $N$ is, the more modes can exist. These can be deduced from the combination of inwards and outwards pointing surface charge distributions in the different arms under the constraint of the rotational symmetry. We will here discuss the example of $N=6$ arms as shown in the right column of Fig.~\ref{fig:BEM}. First of all, we notice that a number of resonances appear, which can be quiet broad. As in the case of the trimer nanoantenna, we see that for $|\ell|=1,2$ and $4$ the resonances split depending on the class of the twisted light beam, while for $|\ell|=3$ again both classes excite the same mode. Looking in more detail, we find that now four modes are excited. Denoting the $\ell=0$ mode at $\lambda_b=780$~nm `bright', we see that in this mode the surface charge distributions in half of the arms are pointing inwards and the ones in the other half are pointing outwards. For $\ell=-1$ we see that a mode at $\lambda_{d1}=615$~nm with all surface charge distributions pointing synchronously inwards is excited in agreement with the radial character of the incident twisted light. Similar to coupled harmonic oscillators, where the out-of-phase oscillation is the energetically highest one, this is the mode with the highest energy. For $\ell=+1$ also a dark mode at $\lambda_{d2}=825$~nm is excited, which is energetically more favourable than the bright mode. Here, the surface charge distribution of two opposing arms point inward, while ones of the other four arms point outward. For $\ell=-2$ the bright mode is excited, but $\ell=+2$ excites a third dark mode at $\lambda_{d3}=830$~nm, in which the surface charge of the arms is alternating. For $\ell= \pm3$, we then see that the dark mode $\lambda_{d1}$ is excited for both parallel and antiparallel classes, while for $|\ell|=4$ the resonances split again. In total, we find four different modes. Note that there is a small shift in the resonances, which can be explained by an increased beam waist for increasing $|\ell|$. 

We have seen that with twisted light a number of dark modes in a nanoantenna composed of several arms can be excited and that the resonance modes are selective on both the absolute value of the OAM $|\ell|$ and on the sign of the OAM $\ell$. We further tested, that the found resonance behaviour is robust against small changes of the position of the beam axis: as long as the beam axis lies within the gap region, the same resonance behaviour is found. This shows that the read out of the value of OAM with such nanoantenna arrays is possible.

\begin{figure}
\includegraphics[width=1\columnwidth]{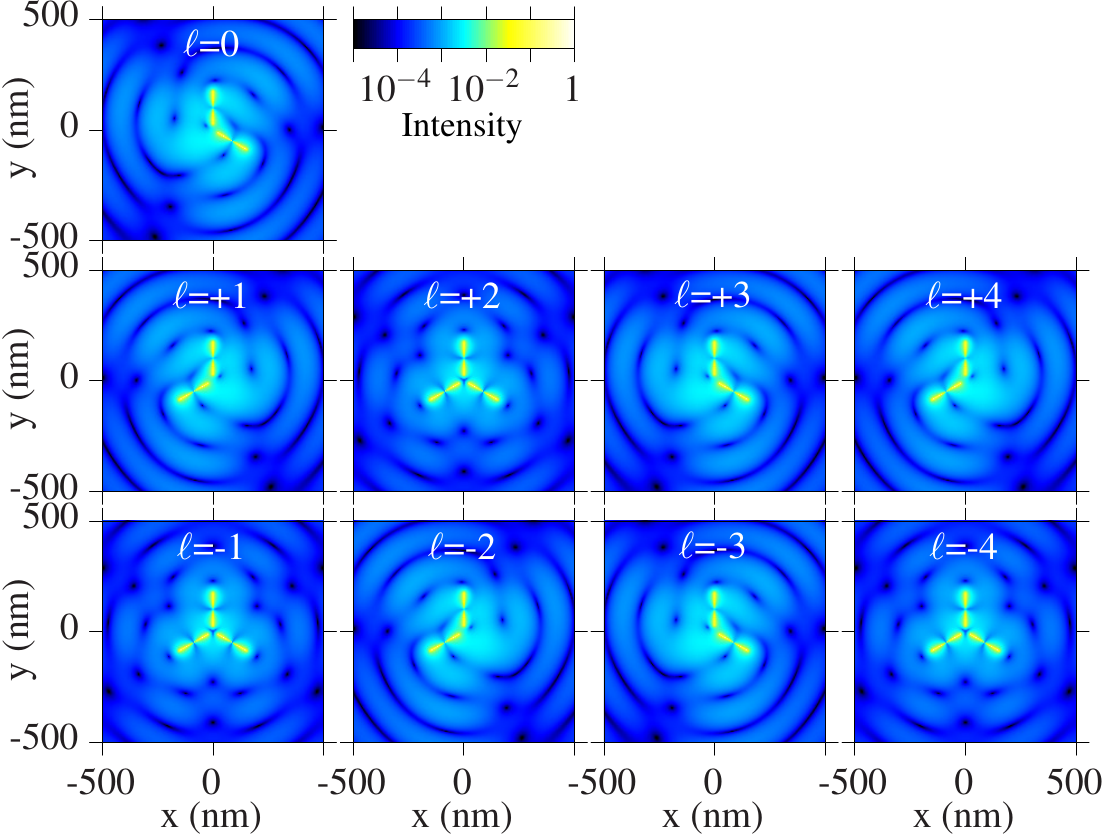}
\caption{Induced electric field calculated with the antenna model. The intensity in the $z=0$ plane with a logarithmic scale is shown for $\ell=0$ (top) and positive OAM $\ell=1\ldots 4$ (middle) and negative OAM $\ell=-1 \ldots -4$ (bottom).
\label{fig:antennamodel}}
\end{figure}

Now we discuss the electric field from the analytical antenna model for the trimer antenna with $N=3$ in Fig. \ref{fig:antennamodel}. Here, the intensity of the electrical field is plotted in the $z=0$ plane with a logarithmic scale. From the field we can distinguish several modes of excitation. For $\ell=0$ and $\ell=\pm 3$ there is the same intensity distribution, where the ends of the middle and right wire show high intensities. In contrast for $\ell=+1,-2,+4$ the ends of the left and middle wire show high intensities. The time dependence of the fields for $\ell=0$ and $\pm 1$ clearly shows that the rotation direction in these two cases is inverted. However, the energy of these two modes is degenerate as they have the same symmetry. A different pattern is found for $\ell=-1,+2,-4$, where all ends of the three wires have high intensity and in the middle of the antenna is a dark spot with low intensity. This is a breathing mode, where only the intensity but not the field pattern changes with time \cite{schmidt2012dark}. The field patterns agree with the resonance behaviour found in the BEM calculations, for the rotating mode for $\ell=0,+1,-2,\pm3,+4$ we find an excitation of the bright mode, while the breathing mode for $\ell=-1,+2,-4$ corresponds to the dark mode. This shows that our analytical model is capable of identifying the different modes excited by the twisted light.

\begin{figure}
\includegraphics[width=1\columnwidth]{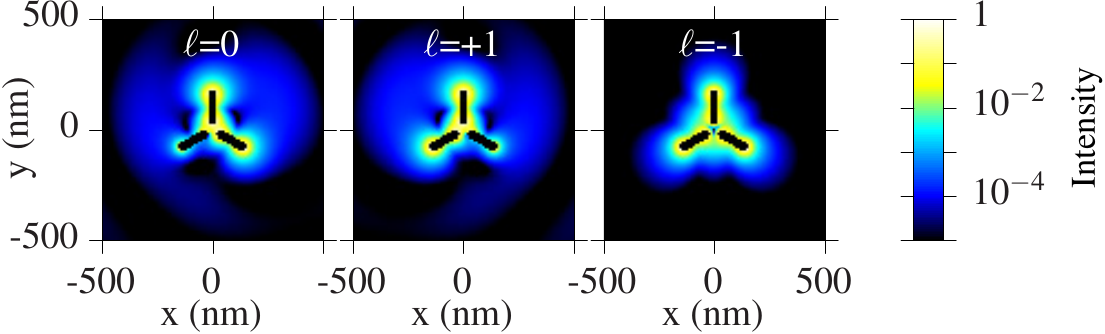}
\caption{Electric field calculated with the BEM . The intensity in the $z=0$ plane is shown on a logarithmic scale for $\ell=0,+1,-1$ from left to right.
\label{fig:BEMintensity}}
\end{figure}

It is interesting to compare the analytical model to the calculations done with the BEM. For this end we plot the electric field intensity calculated with the BEM in Fig. \ref{fig:BEMintensity} for the cases $\ell=0,\pm1$, where the black areas in the middle of the figures show the positions of rods. For the wavelength, we choose the resonant wavelength $\lambda_b$ or $\lambda_d$. Again we find rotating modes for $\ell=0$ and $\ell=+1$, which differs only in the rotation direction in the time behaviour. These fields correspond to the bright mode. For $\ell=-1$ we also find a breathing behaviour as expected for the dark mode.
\begin{figure}
	\includegraphics[width=1\columnwidth]{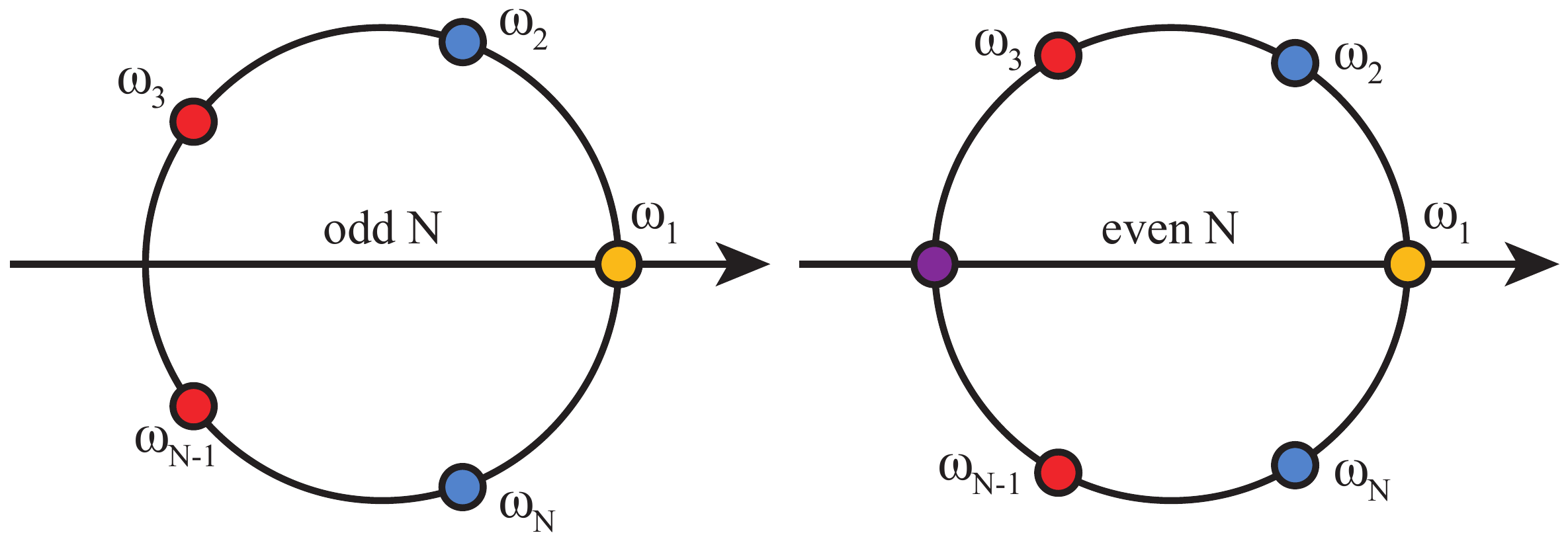}
	\caption{Sketch of the possible resonance frequencies and their degeneration for and odd (left) and even (right) number of arms. Same colours imply that the modes are degenerate. 
	\label{fig:degeneration}}
\end{figure}

Having understood the resonance behaviour in detail, we now summarize the resonance behaviour using symmetry arguments only. For instance, in the case of three antenna arms every antenna arm can have a different phase of the radiating field. The first possibility, denoted by $\omega_1$, is that all arms have the same phase $\omega_1$:$(0,0,0)$. The second possibility is that the second antenna arm has an additional phase of $2\pi/3$ and the third one of $4\pi/3$ so that $\omega_2$:$(0,2\pi/3,4\pi/3)$. The third option is that there is a phase change by $4\pi/3$ of every antenna arm $\omega_3$:$(0,4\pi/3,8\pi/3)$. By reducing the phase changes into a $2\pi$-interval, it is clear that the latter modes have equal frequencies $\omega_2=\omega_3$ with their phases in an inverted order. This explains the excitation of the degenerate modes with a different temporal behaviour. These considerations can be generalized for $N$ antenna arms, where we have $N$ possible modes with:
\begin{eqnarray}
\omega_i &:&(0,1,2,\ldots,N-1) \times \frac{(i-1)\cdot 2\pi}{N}
\end{eqnarray}
where $i= 1,2,\ldots,N$.
The second and the $N$-th option are degenerate $\omega_2=\omega_N$, the third and the $(N-1)$-th $\omega_3=\omega_{N-1}$ and so on. In total for an odd number of antenna arms there are $(N-1)/2$ degenerate modes and $(N-2)/2$ for an even number of antenna arms. Therefore the number of different resonance frequencies excited by twisted light is $[N/2]+1$ where $[]$ is the Gauss floor function. This is illustrated in the sketch in Fig. \ref{fig:degeneration}. Going back to our discussion of the case of $N=6$, where we found four modes in the BEM, we find that this is consistent with our explanation using the symmetry arguments.

\section{Conclusion}

In conclusion, we have demonstrated that a rotationally arranged nanoantenna can be used to convert the phase information of a twisted light beam into spectral information and hence can be used to classify the phase state of a twisted light beam. To be specific, for a dimer antenna we can distinguish between even and odd values of OAM, while for an antenna with $N\ge 3$ arms also the class of the OAM beam can be determined. We further showed that different dark modes can be excited by a twisted light beam, which cannot be easily excited by plane wave light. Our findings are well explained by a thin-wire model, which gives a handy rule to determine the number of resonance frequency of an antenna consisting of $N$ arms excited by twisted light. Our work shows that also on the nanoscale the OAM influences strongly the optical properties of plasmonic particles and our results will be useful for implementing twisted light beams on the nanoscale for secure communication \cite{ren2016chip,mirhosseini2015high}.

\section{Acknowledgements}

We gratefully acknowledge the Matlab BEM package provided by A. Tr\"ugler and U. Hohenester. D.E.R. is grateful for financial support from the German Academic Exchange Service (DAAD) within the P.R.I.M.E. program. J.M.F. acknowledges financial support from UK's Engineering and Physical Sciences Research Council (grant number: 1580548). S.S.O. and O.H. acknowledge financial support from the Leverhulme Trust (RPG-2014-068). 
We are grateful for fruitful discussion with Rudi Bratschitsch and colleagues as well as with Tilmann Kuhn.


\providecommand{\latin}[1]{#1}
\providecommand*\mcitethebibliography{\thebibliography}
\csname @ifundefined\endcsname{endmcitethebibliography}
  {\let\endmcitethebibliography\endthebibliography}{}

\end{document}